\documentclass[a4paper,10pt,twoside]{cpc-hepnp}
\usepackage{CJK,upgreek,fancyhdr}
\usepackage{multicol}
\usepackage{graphicx}
\usepackage{booktabs}
\usepackage{amssymb,bm,mathrsfs,bbm,amscd}
\usepackage[tbtags]{amsmath}
\usepackage{lastpage}
 \usepackage{multirow}
\usepackage[colorlinks]{hyperref}

\begin{document}
\begin{CJK*}{GB}{gbsn}

\fancyhead[c]{\small }
\fancyfoot[C]{\small \thepage}

\footnotetext[0]{}

\title{Resonant Production of Color Octet Muons at the Future Circular Collider
Based Muon-Proton Colliders\thanks{ This study is supported by TUBITAK under the grant no 114F337.}}

\author{%
     Y. C. Acar $^{1)}$\email{ycacar@etu.edu.tr}%
\quad U. Kaya $^{1, 2)}$\email{ukaya@etu.edu.tr}%
\quad B. B. Oner$^{1}$\email{b.oner@etu.edu.tr}%
\quad
}
\maketitle

\address{%
$^1$ Department of Material Science and Nanotechnology, TOBB University of Economics and Technology, Ankara 06560, Turkey\\
$^2$ Department of Physics, Ankara University,  Ankara 06560, Turkey\\
}

\begin{abstract}
We investigate the resonant production of color octet muons in order
to explore the discovery potential of the FCC-based $\mu p$ colliders.
It is shown that search potential of $\mu p$ colliders essentially
surpass the potential of the LHC and would exceed that of the FCC $pp$ collider.
\end{abstract}

\bigbreak


\footnotetext[0]{\hspace*{-3mm}\raisebox{0.3ex}{}
}%

\begin{multicols}{2}

\section{Introduction}

High energy physics experiments performed in recent decades show
that Standard Model (SM) is consistent in a low energy regime. However,
there are still phenomenological and theoretical problems and questions
to be answered. Experimental research for the new physics, searching for these answers for higher energies, relies on recently developed accelerator
technologies. Energy frontier lepton colliders seem to be prominent
candidates to investigate the validity of the SM at high energies, and they
have the potential to reveal novelties that lie beyond the Standard Model
(BSM). Producing and colliding muon beams with intense bunches to
achieve sufficiently high luminosities is still a promising topic.
In this regard, a recent paper by the Muon Accelerator Program (MAP)
addressed designs of various center of mass (CM) energy muon colliders
($\mu$C) from 126 GeV (Higgs-factory) to multi-TeV (energy frontier)
options \cite{lab1}. Also ultimate case muon colliders with CM energy
up to 100 TeV were considered in another study, and parameters of these
colliders were given \cite{lab2}. 

Developing the technology of lepton colliders makes high luminosity and
high CM energy lepton-hadron colliders possible. In this manner, one
can utilize the advantages of their vital role in understanding the fundamental
structure of matter using the highest energy hadron beams, which will
be provided by the Future Circular Collider (FCC) \cite{lab10}.
In the near future, it is expected that the construction of $\mu p$ machines
can also be considered, depending on the solutions to the principal issues
of the $\mu^{+}\mu^{-}$ colliders.  Some advantages of the highest energy $\mu p$ machines can be listed briefly as follows. Firstly, multi-TeV scale muon-proton collisions are testing mechanisms of composite models and may give us clear hints about the fermion mixing and generation replication puzzle of the SM fermions. In addition, they would present experimental results that enable us to understand the QCD better. Exotic particle productions are more probable compared to the ep colliders since the large mass ratio between muon and electron \cite{lab3}.

Muon-proton colliders were proposed two decades ago. Construction
of additional proton ring in $\sqrt{s}$= 4 TeV muon collider tunnel
was suggested in \cite{lab3} to handle $\mu p$ collider with the
same CM energy. However, luminosity value, namely $L_{\mu p}=3\times10^{35}cm^{-2}s^{-1}$,
was extremely overestimated, a realistic value for this should
be three orders smaller \cite{lab4}. Then, construction of additional
200 GeV energy muon ring in the Tevatron tunnel in order to handle
$\sqrt{s}$ = 0.9 TeV $\mu p$ collider with $L_{\mu p}=10^{32}cm^{-2}s^{-1}$
was considered in \cite{lab5}. Also in Ref. \cite{lab4} the ultimate
case of muon beams with 50 TeV energy \cite{lab2} had been taken into account
as an option for 100 TeV CM energy $\mu p$ colliders assuming that a
50 TeV proton ring would be added into the $\mu$C tunnel and a luminosity
value $\sim10^{33}\:cm^{-2}s^{-1}$ is estimated. The FCC based muon-proton
and muon-lead ion colliders' main parameter calculations were performed
in a recent paper which considers beam-beam effects and a basic collider
parameter optimization \cite{lab6}. 

In Ref. \cite{lab7}, the physics potentials of $\mu p$ colliders with
several energy and luminosity options (from $\sqrt{s}=314$ GeV, $L_{\mu p}=0.1\:fb^{-1}$
per year to $\sqrt{s}=4899$ GeV, $L_{\mu p}=280\:fb^{-1}$ per year)
were studied. The sensitivity reach of each collider was calculated for
some BSM phenomena such as R-parity violating squarks, leptoquarks,
leptogluons and extra-dimensions. Similarly, R-parity violating resonances
were examined for Tevatron based $\mu p$ collider with $\sqrt{s}=0.9$
TeV and $L_{\mu p}=10^{32}cm^{-2}s^{-1}$ in \cite{lab8}. In a recent
study excited muon production was analyzed at muon-hadron colliders
based on the FCC \cite{lab9}. 

This paper shows a follow up work of our previous study which was
based on the search potential of the FCC-based ep colliders on color
octet electrons \cite{lab11} (besides, there are number of papers
devoted to the study of color octet electron production at the LHC
\cite{lab1212,lab1313,lab1414,lab1515} and LHeC \cite{lab1616,lab1717,lab1818}).

We now consider another design, namely, the construction of a muon ring
tangential to the FCC, which is schematically shown in Fig. 1. The aim
is to achieve the highest possible CM energies in lepton-hadron colliders
in order to make some of the BSM physics research possible.
Here, the physics potential of these future colliders is revealed by quantitatively
exploring resonant production of color octet muons. Parameters of
the FCC-based muon proton colliders are given in Table I. The first
four colliders \cite{lab6} use the most recent design parameters
of MAP \cite{lab1}. The last row corresponds to the ultimate case
with 20 TeV muon beams in the FCC tunnel. 20 TeV choice is due to synchrotron
radiation loss of muons which is desired to be limited at 1 GeV/turn
for a muon accelerator with 100 km circumference \cite{lab12}.

\begin{center}
\includegraphics[width=8cm]{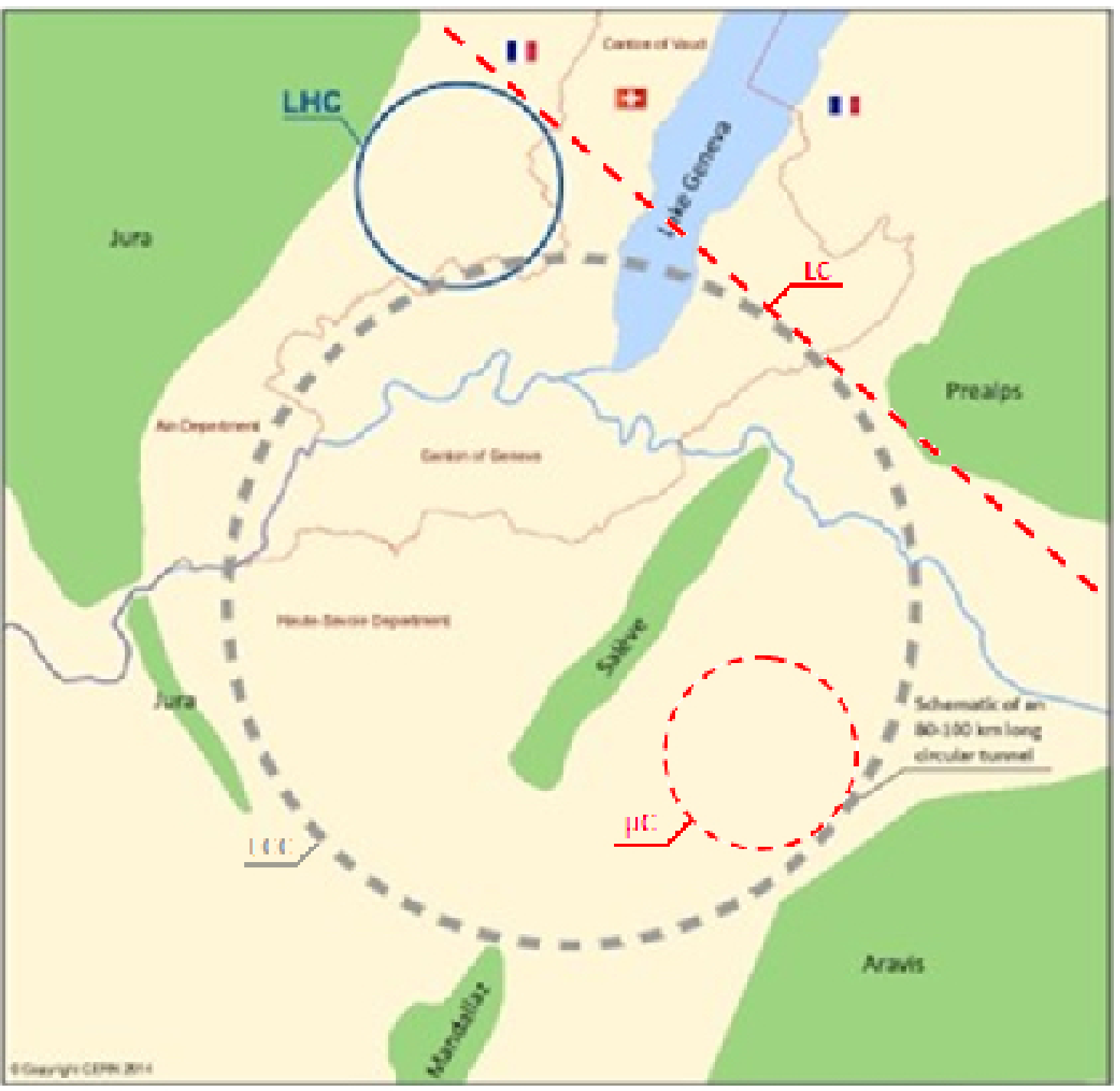}
\figcaption{\label{fig1}   Possible configuration of the FCC, linear collider
(LC) and muon collider ($\mu$C). }
\end{center}

\begin{center}
\columnbreak
\tabcaption{ \label{tab1}  Main parameters of the FCC based $\mu p$ colliders.}
\footnotesize
\begin{tabular*}{80mm}{c@{\extracolsep{\fill}}ccc}
\toprule Collider & $E_{\mu}$& CM Energy& $L_{int}$, $fb^{-1}$\\
              Name&        (TeV)&                     (TeV)&                     (per year) \\
\hline

{$\mu$63$\otimes$FCC} & 0.063 & 3.55 & 0.02\\
{$\mu$750$\otimes$FCC} & 0.75 & 12.2 & 5\\
{$\mu$1500$\otimes$FCC} & 1.5 & 17.3 & 5\\
{$\mu$3000$\otimes$FCC} & 3.0 & 24.5 & 5\\
{$\mu$20000$\otimes$FCC} & 20 & 63.2 & 10\\
\bottomrule
\end{tabular*}
\vspace{0mm}
\end{center}
\vspace{0mm}

The rest of the paper is organized as follows. In Section II, we present
phenomenology of color octet muon. Section III covers signal-background
analyses and is closed by giving the results of discovery limit searches
of muon-proton colliders. Section IV addresses the determination of
compositeness scales via muon-proton collider options under two possible cases regarding the results of the FCC. Finally, Section V contains summary
of the obtained results. 

\section{Color octet muon}

One of the possible answer to the problems mentioned
in the Introduction may hide behind the concept of compositeness.
Fermion-scalar and three-fermion models are the most proper options
which enable known SM leptons to be constructed from more fundamental
particles, namely preons. If the SM leptons are composed of color triplet
fermions and color triplet scalars, then both fermion-scalar and three-fermion
models predict at least one color octet partner to the color singlet
leptons:

\begin{eqnarray}
\label{eq1}
\ensuremath{\ell}=(F\bar{S})=3\otimes\bar{3}=1\oplus8
\end{eqnarray}

\begin{eqnarray}
\label{eq2}
\ell=(FFF)=3\otimes3\otimes3=1\oplus8\oplus8\oplus10.
\end{eqnarray}

Interaction lagrangian of $\ell_{8}$ with leptons and gluons can
be written as 

\begin{eqnarray}
\label{eq3}
L=\frac{1}{2\Lambda}\sum_{l}\left\{ \bar{\ell_{8}^{\alpha}}g_{s}G_{\mu\nu}^{\alpha}\sigma^{\mu\nu}\left(\eta_{L}\ell_{L}+\eta_{R}\ell_{R}\right)+h.c.\right\}
\end{eqnarray}

\noindent where $g_{s}$ is strong coupling constant, $\varLambda$ denotes
compositeness scale, $G_{\mu\nu}$ is gluon field strength tensor,
$\ell_{L(R)}$ stands for left (right) spinor components of lepton,
$\ell=e,\:\mu,\:\tau$; $\sigma^{\mu\nu}$ is the antisymmetric tensor
($\sigma^{\mu\nu}=\frac{i}{2}\left[\gamma^{\mu},\:\gamma^{\nu}\right]$),
$\eta_{L}(\eta_{R})$ symbolizes chirality factor. Keeping in mind
leptonic chiral invariance ($\eta_{L}\eta_{R}=0$), we take $\eta_{L}=1$
and $\eta_{R}=0$. Decay width of $\ell_{8}$  is given by 

\begin{eqnarray}
\label{eq4}
\Gamma(\ell_{8}\rightarrow\ell+g)=\frac{\alpha_{s}M_{\ell_{8}}^{3}}{4\Lambda^{2}}
\end{eqnarray}

\noindent where $\alpha_{s}=g_{s}/4\pi$. Dependence
of the decay width on the mass of $\mu_{8}$ is presented in Fig.
2 for $\varLambda=M_{\mu_{8}}$ and $\varLambda=100$ TeV cases.

\end{multicols}

\begin{center}
\includegraphics[width=18cm]{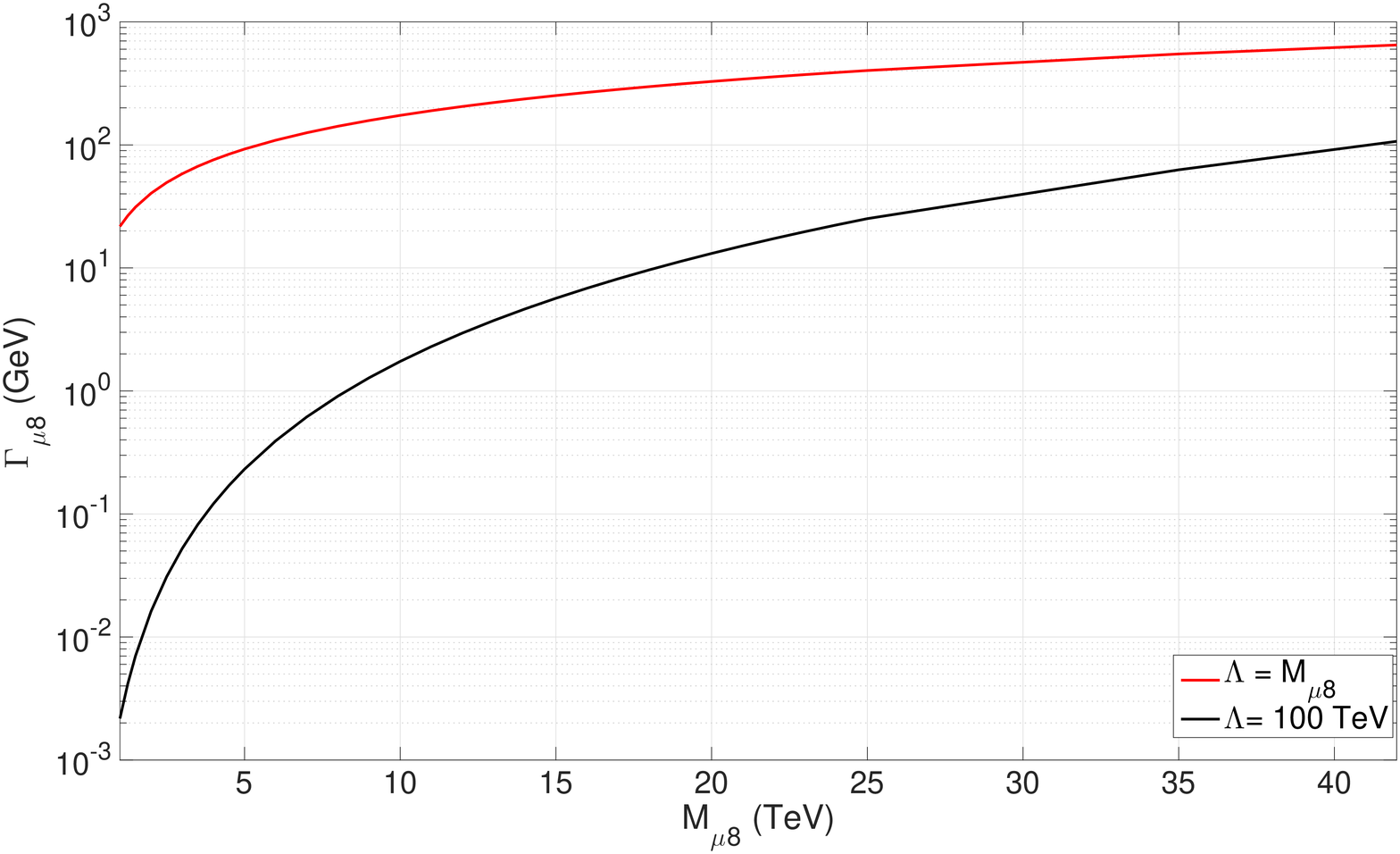}
\figcaption{\label{fig2} Color octet muon decay width for $\Lambda=M_{\mu8}$ and $\Lambda=$100
TeV. }
\end{center}

\begin{multicols}{2}

The resonant $\mu_{8}$ production (see Figure
3) cross sections for different stages of the FCC based $\mu p$ colliders
from Table I were calculated using MadGraph5 event generator \cite{lab13}.
CTEQ6L1 parametrization \cite{lab14} was used as parton distribution
function and results were presented in Figure 4. MadGraph5-Pythia6
interface was used for parton showering and hadronization \cite{lab15}. The same tools were used for the rest of the study and further calculations did not take detector effects into account.

\begin{center}
\includegraphics[width=6cm]{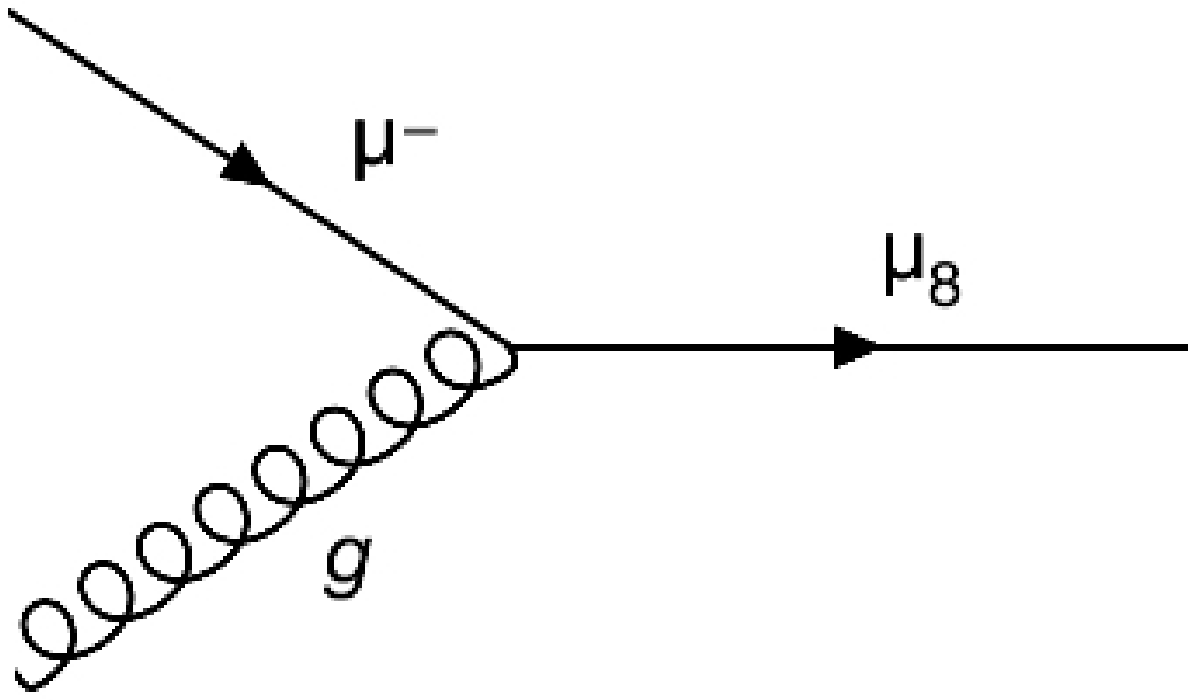}
\figcaption{\label{fig3}   Feynman diagram for the resonant $\mu_{8}$ production. }
\end{center}

\end{multicols}
\begin{center}
\includegraphics[width=18cm]{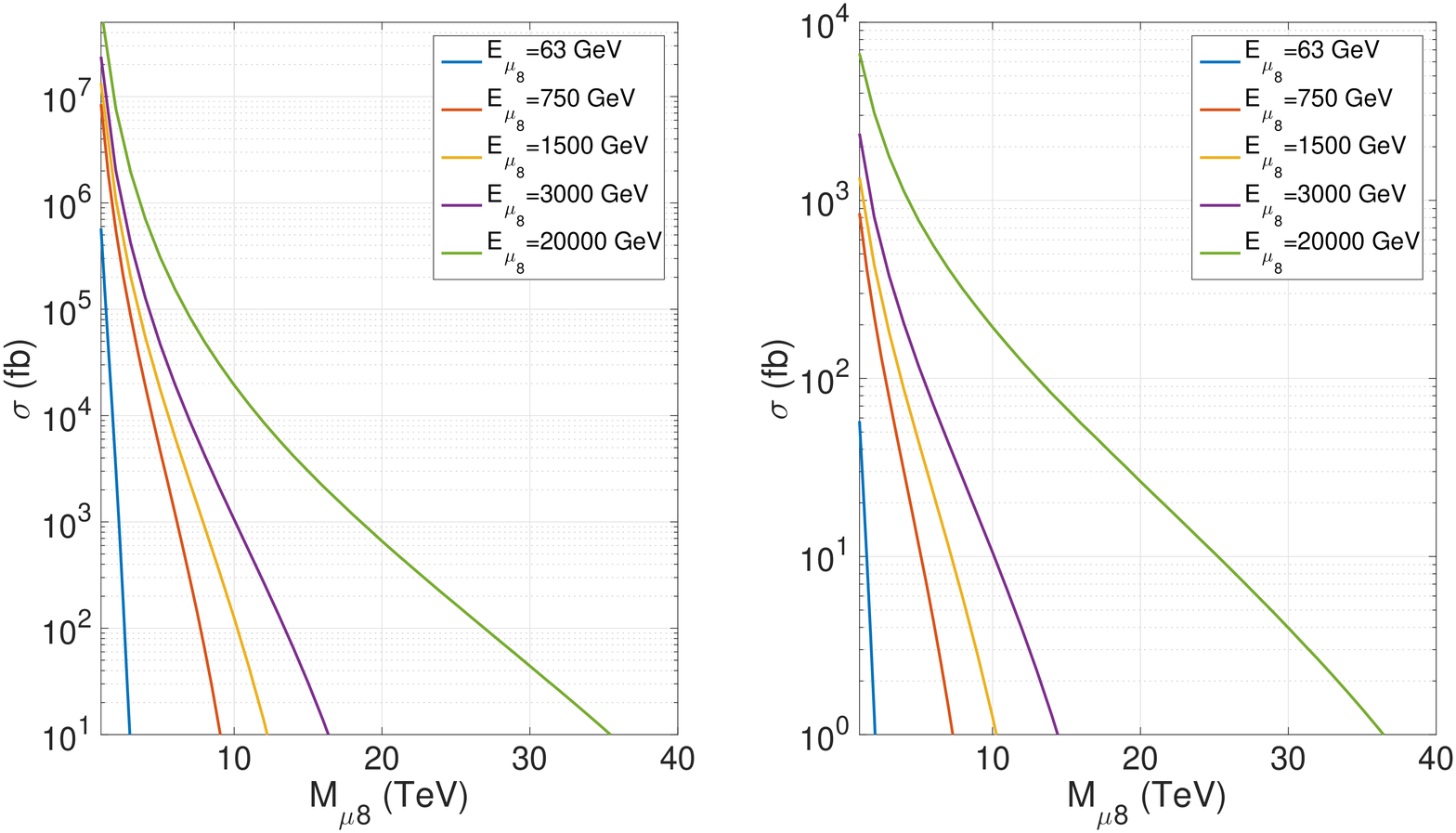}
\figcaption{\label{fig4} Resonant color octet muon production at the FCC based $\mu p$ colliders
given in Table I for (a) $\Lambda=M_{\mu8}$ and for (b) $\Lambda=$100
TeV ($\mu p \rightarrow \mu_8 X \rightarrow \mu j X$). }
\end{center}
\begin{multicols}{2}

\section{Signal - Background Analysis}

In this section, results of the numerical calculations are shown for
the process $p\mu\rightarrow j\mu$ to leading order in order to analyze the search potential
of the FCC based muon-proton colliders on the $\mu_{8}$ discovery
via resonant production within $\Lambda=M_{\mu8}$ scenario. Let us mention that jet corresponds to gluon for signal ($\mu g \rightarrow \mu_{8}  \rightarrow  \mu g$ at partonic level) and quarks for main background ($\mu q \rightarrow \mu q$ through $\gamma$ and Z exchanges) processes.

A staged approach was applied to determine mass limits as follows.
$\mu$63$\otimes$FCC, the $\mu p$ collider
with minimum CM energy, was chosen as the initial collider where discovery
limit of $\mu_{8}$ mass was to be sought. After the discovery limit was
determined, a worse scenario was considered where $\mu_{8}$ was assumed
to have a larger mass. It was supposed that the previous collider had excluded $\mu_{8}$ mass up to the
corresponding discovery limit and necessary cuts regarding
this assumption were applied for the next higher CM energy $\mu p$ collider. Latter colliders follow the rows of Table I respectively.
This procedure ends up with the ultimate $\mu p$ collider with CM energy
63.2 TeV which was given in the last row of the Table I. One should note that a sequential building of these colliders seems not to be the realistic case. Therefore, if a muon-proton collider is built, color-octet muon search discovery cuts would depend on up-to-date experimental exclusion limits.

\end{multicols}
\begin{center}
\includegraphics[width=18cm]{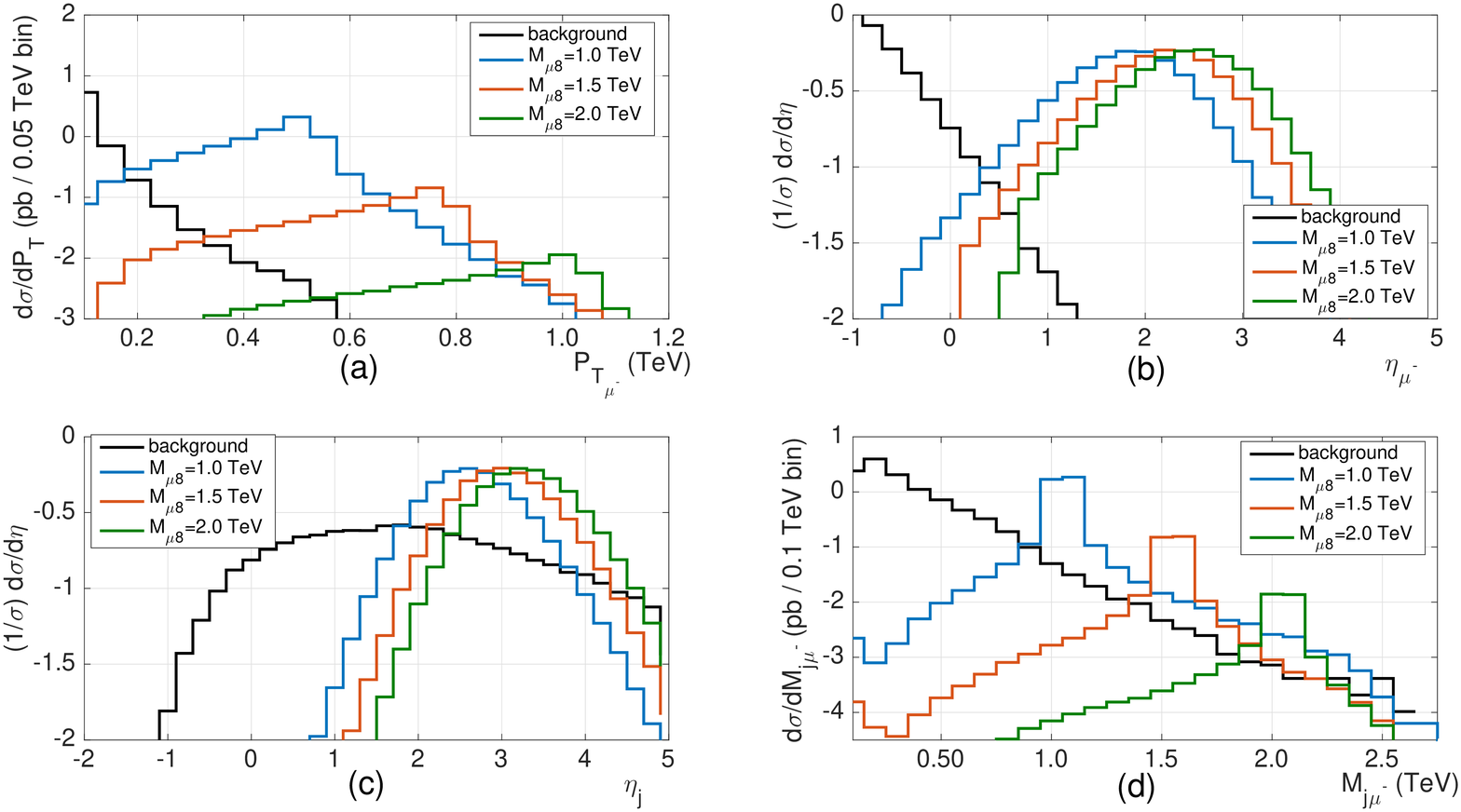}
\figcaption{\label{fig5} a) Transverse momentum distributions of final state muons (almost same distribution holds for the leading-jets),
b) pseudorapidity distributions of final state muons, c) pseudorapidity
distributions of final state jets and d) invariant mass distributions
for signal and background at $\mu63\otimes$FCC after generic cuts. }
\end{center}
\begin{multicols}{2}

Kinematical distributions of $\mu$63$\otimes$FCC
with generic cuts ($p_{T_{\mu}}>20$ GeV, $p_{T_{j}}>30$ GeV) are
given in Fig. 5. Reconsideration of the ATLAS/CMS results in the search
for the second generation leptoquarks \cite{lab16,lab17} (which
have the same decay channel as $\mu_{8}$) leads us to the strongest current
limit on the color octet muon mass, $M_{\mu_{8}}\gtrsim1$ TeV. Therefore,
we chose the discovery cut for transverse momentum to be $p_{T}>350$ GeV
on our initial $\mu p$ collider. This transverse momentum cut was applied on final state muon as well as leading-jet. In order to suppress the background
while keeping the signal cross section as much as possible, the following
pseudorapidity cuts were also applied: $2.00<\eta_{j}<4.00$ , $0.5<\eta_{\mu}<4.74$.
Maximum possible value of $\eta_{\mu}$ and $\eta_{j}$ was taken 4.74
which corresponds to $1^{o}$ in proton direction. This value can
be covered by very forward detector as in the LHeC case \cite{lab18}.
Effects of these discovery cuts can be seen by comparing Fig. 5(d) with Fig. 6 where invariant mass
of $\mu_{8}$ was reconstructed from final state particles $\mu$ and
leading-jet. After these cuts, cross sections of signals remained almost the same while background cross-section decreased remarkably. 

\begin{center}
\includegraphics[width=10cm]{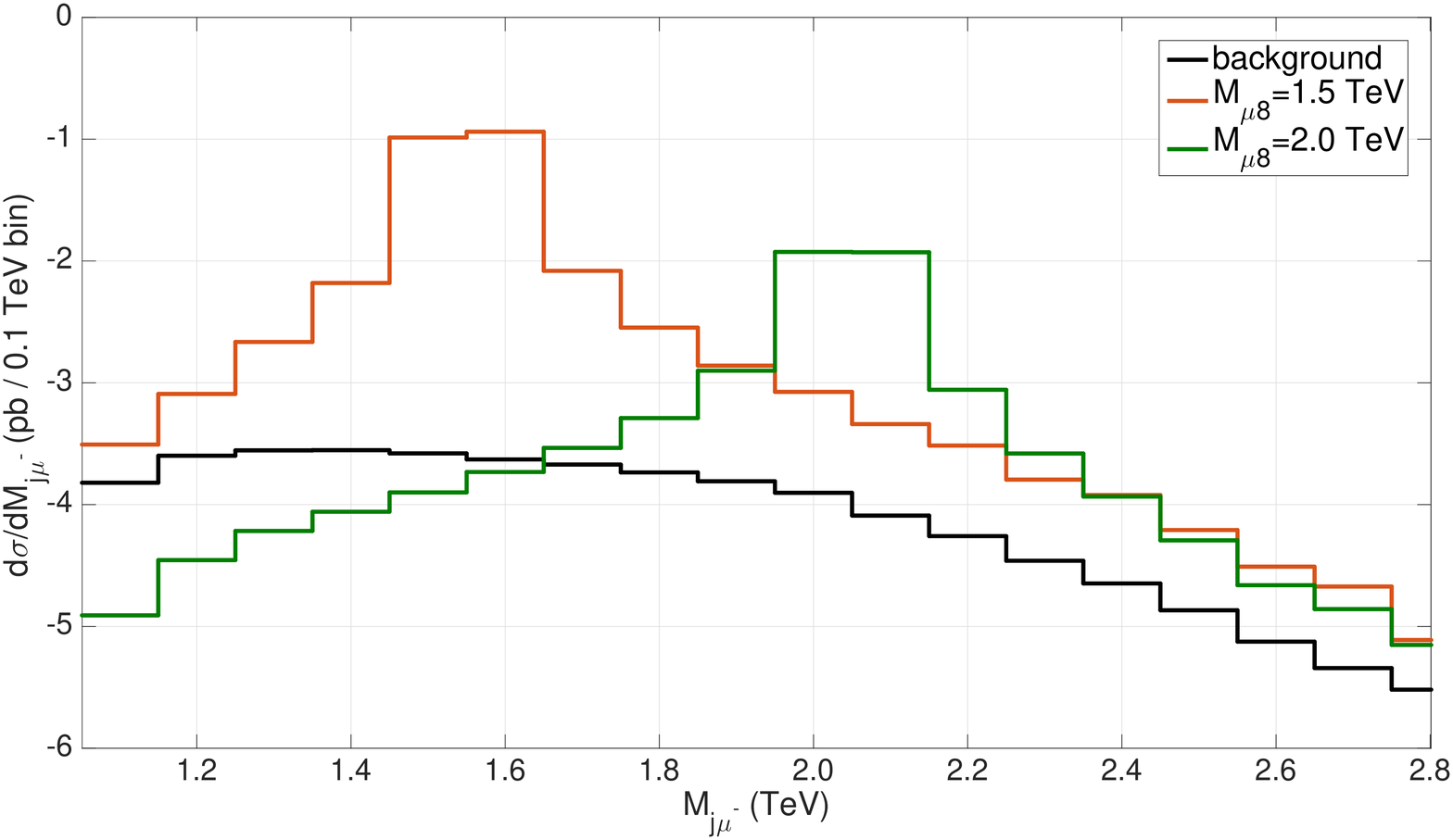}
\figcaption{\label{fig6}   Invariant mass distributions for signal and background at $\mu63\otimes$FCC
after discovery cuts. }
\end{center}

Statistical significance ($SS$) is calculated using the formula below: 
\begin{eqnarray}
SS=\sqrt{2\:L_{int}}\sqrt{(\sigma_{S}+\sigma_{B})\,ln(1+(\sigma_{S}/\sigma_{B}))-\sigma_{S}}\label{eq5}
\end{eqnarray}

\noindent where $\sigma_{S}$ and $\sigma_{B}$ denote cross-section
values of signal and background, respectively. Integrated luminosity
values, $L_{int}$, of each collider per year was estimated in \cite{lab6}.
Discovery ($SS=5$) and observation ($SS=3$) limits for 0.02 $fb^{-1}$
$\mu$63$\otimes$FCC integrated luminosity
were found to be 2380 and 2460 GeV, respectively. Regarding these results
of the minimum energy $\mu p$ collider, $p_{T}>800$ GeV was considered
appropriate for the next stage $\mu$750$\otimes$FCC
and similar analyses were performed. These consecutive calculations
gave us mass reach of each collider as given in Table II. Applied
discovery cuts were also given in the same table and mass window formulation
was kept same for all calculations: $M_{\mu8}-2\Gamma_{\mu8}<M_{\mu8}<M_{\mu8}+2\Gamma_{\mu8}$. Signal and background event numbers were calculated directly in this mass window without using any binning algorithm.
Invariant mass distributions after discovery cuts related to higher
energy colliders are presented in Figure 7.

\end{multicols}

\begin{center}
\includegraphics[width=18cm]{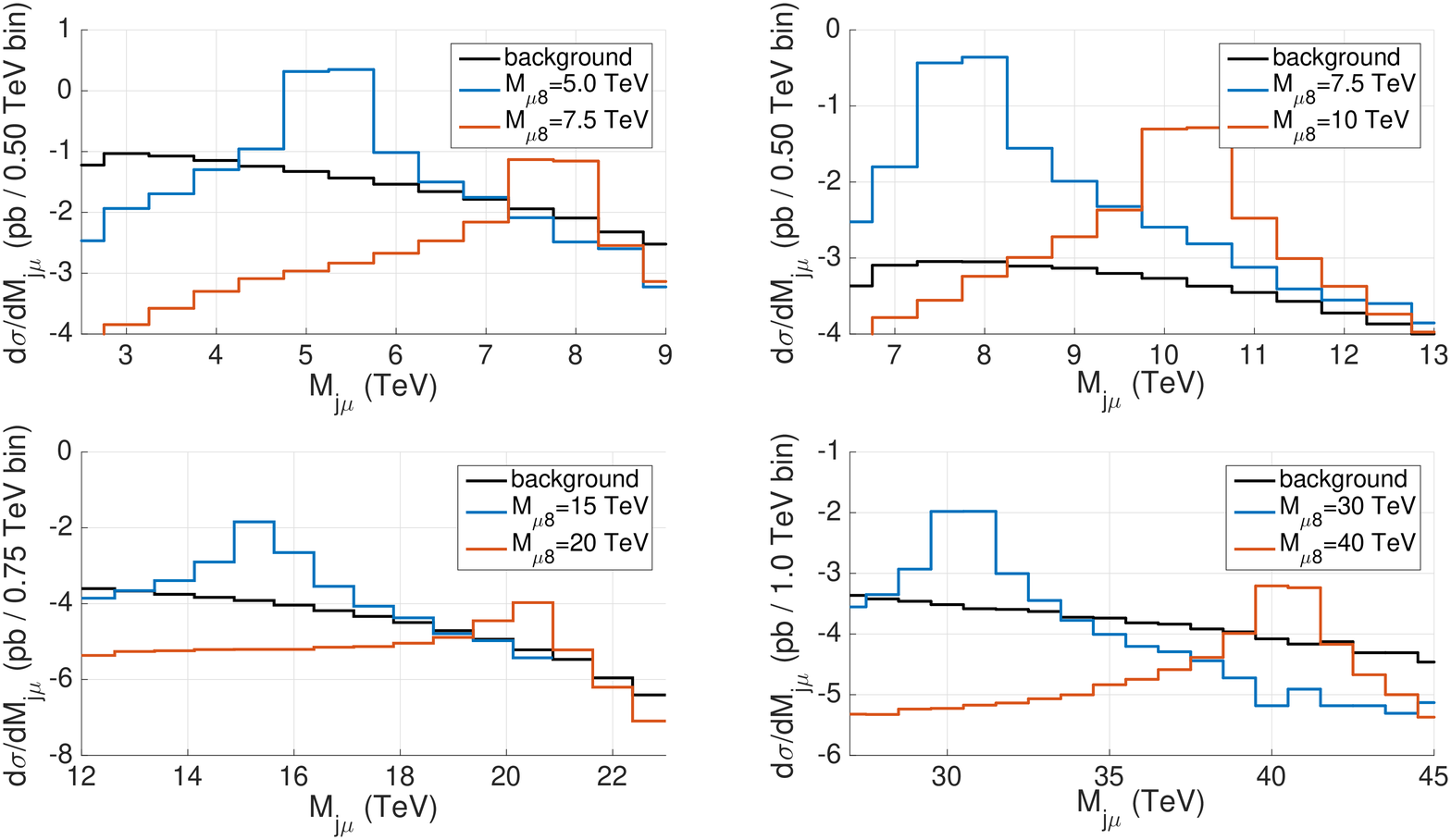}
\figcaption{\label{fig77} Invariant mass distributions for signal and background at a) $\mu750\otimes$FCC,
b) $\mu1500\otimes$FCC, c) $\mu3000\otimes$FCC and for d) the ultimate
case $\mu20000\otimes$FCC colliders after discovery cuts. }
\end{center}

\begin{multicols}{2}

\end{multicols}

\begin{table}[]
\centering{}%
\caption{\label{tab2}  Kinematical discovery cuts and observation ($3\sigma$) and discovery
($5\sigma$) limits for $\mu_{8}$ at different $\mu p$ colliders. Transverse momentum cuts are given in TeV. Significance values are calculated locally.}
\label{tab2}
\begin{tabular}{lclclclclclclclclclcl}
\hline
 \multicolumn{1}{|c}{ \multirow{2}{*}{Collider Name} }& \multicolumn{1}{|c|}{\multirow{2}{*}{$L_{int}$, $fb^{-1}$}} & \multicolumn{5}{c|}{Kinematical Cuts}                                                                             & \multicolumn{4}{c|}{$M_{\mu8}$ {\small $\pm PDF\% $} $\pm scale\%$, TeV}                         \\ \cline{3-11} 
               \multicolumn{1}{|c}{}     &      \multicolumn{1}{|c|}{}             &   \multicolumn{1}{c|}        {$p_{T_{min}}$}        &        \multicolumn{1}{c|}        { $\eta_{\mu_{min}}$}           &          \multicolumn{1}{c|}        {$\eta_{\mu_{max}}$}         &             \multicolumn{1}{c|}        { $\eta_{j_{min}}$}       &     \multicolumn{1}{c|}        { $\eta_{j_{max}}$}              &\multicolumn{2}{c|}{ $3\sigma$} & \multicolumn{2}{c|}{$5\sigma$} \\ \hline

 \multicolumn{1}{|c}{ \multirow{2}{*}{$\mu$63$\otimes$FCC}} & \multicolumn{1}{|c|}{ \multirow{2}{*}{0.02}} & \multicolumn{1}{c|}{\multirow{2}{*}{0.350}} & \multicolumn{1}{c|}{\multirow{2}{*}{0.5}} & \multicolumn{1}{c|}{\multirow{2}{*}{4.74}} & \multicolumn{1}{c|}{\multirow{2}{*}{2.0}} & \multicolumn{1}{c|}{\multirow{2}{*}{4.0} }& \multirow{2}{*}{2.46}  &  {\small +2.60\% +1.63\%}  &\multicolumn{1}{|c}{ \multirow{2}{*}{2.38}}  & \multicolumn{1}{c|}   {\small +2.52\% +1.68\%}  \\ \cline{9-9} \cline{11-11} 
        \multicolumn{1}{|c}{}              &               \multicolumn{1}{|c|}{}    &          \multicolumn{1}{c|}{}          &            \multicolumn{1}{c|}{}        &          \multicolumn{1}{c|}{}          &          \multicolumn{1}{c|}{}          &           \multicolumn{1}{c|}{}         &             \multicolumn{1}{|}{}       &{\small -1.83\% -1.75\%}  &        \multicolumn{1}{|c}{}         & \multicolumn{1}{c|}  {\small -2.10\% -1.89\%}   \\ \hline

 \multicolumn{1}{|c|}{\multirow{2}{*}{$\mu$750$\otimes$FCC}} & \multicolumn{1}{|c|}{\multirow{2}{*}{5}} & \multicolumn{1}{|c|}{\multirow{2}{*}{0.800}} & \multirow{2}{*}{-1.3} &  \multicolumn{1}{|c|}{\multirow{2}{*}{4.74}} & \multirow{2}{*}{1.0} &  \multicolumn{1}{|c|}{ \multirow{2}{*}{4.1}} & \multirow{2}{*}{9.60}  & {\small +1.34\% +0.63\%} & \multicolumn{1}{|c}{\multirow{2}{*}{9.21}}  &\multicolumn{1}{c|}  {\small +1.46\% +0.74\%} \\ \cline{9-9} \cline{11-11} 
  \multicolumn{1}{|c|}{}                 &                   &          \multicolumn{1}{|c|}{}     &                   &          \multicolumn{1}{|c|}{}             &                   &           \multicolumn{1}{|c|}{}            &               &{\small -1.15\% -1.27\%}  &             \multicolumn{1}{|c}{}            &\multicolumn{1}{c|}  {\small -1.20\% -1.37\%} \\ \hline

 \multicolumn{1}{|c|}{\multirow{2}{*}{$\mu$1500$\otimes$FCC}} &\multicolumn{1}{|c|}{ \multirow{2}{*}{5}} &  \multicolumn{1}{|c|}{ \multirow{2}{*}{3.00}} & \multirow{2}{*}{-1.7} & \multicolumn{1}{|c|}{\multirow{2}{*}{4.74}}  & \multirow{2}{*}{0.7} &   \multicolumn{1}{|c|}{\multirow{2}{*}{3.9}} & \multirow{2}{*}{13.8}  &{\small +1.30\% +0.51\%}  & \multicolumn{1}{|c}{\multirow{2}{*}{13.2}}  &\multicolumn{1}{c|} {\small + 1.36\% +0.53\%}  \\ \cline{9-9} \cline{11-11} 
   \multicolumn{1}{|c|}{}                &                   &         \multicolumn{1}{|c|}{}             &                   &        \multicolumn{1}{|c|}{}               &                   &       \multicolumn{1}{|c|}{}                &                    & {\small -1.01\% -0.87\%} &             \multicolumn{1}{|c}{}           &\multicolumn{1}{c|} {\small -1.14\% -1.06\%}  \\ \hline

 \multicolumn{1}{|c|}{\multirow{2}{*}{$\mu$3000$\otimes$FCC}} &\multicolumn{1}{|c|}{ \multirow{2}{*}{5}} &  \multicolumn{1}{|c|}{ \multirow{2}{*}{4.40}} & \multirow{2}{*}{-2.1} & \multicolumn{1}{|c|}{\multirow{2}{*}{4.74}}  & \multirow{2}{*}{0.3} &   \multicolumn{1}{|c|}{\multirow{2}{*}{3.5}} & \multirow{2}{*}{18.9}  &{\small +1.22\% +0.53\%}  & \multicolumn{1}{|c}{\multirow{2}{*}{18.1}}  &\multicolumn{1}{c|} {\small  -1.01\% -0.63\%} \\ \cline{9-9} \cline{11-11} 
  \multicolumn{1}{|c|}{}                 &                   &          \multicolumn{1}{|c|}{}            &                   &           \multicolumn{1}{|c|}{}            &                   &        \multicolumn{1}{|c|}{}               &                    &{\small +1.27\% +0.44\%}  &        \multicolumn{1}{|c}{}                &\multicolumn{1}{c|}  {\small -1.22\% -0.77\%} \\ \hline

 \multicolumn{1}{|c|}{\multirow{2}{*}{$\mu$20000$\otimes$FCC}} & \multicolumn{1}{|c|}{\multirow{2}{*}{10}} &   \multicolumn{1}{|c|}{\multirow{2}{*}{6.00}} & \multirow{2}{*}{-2.7} & \multicolumn{1}{|c|}{\multirow{2}{*}{4.74}}  & \multirow{2}{*}{-0.7} &   \multicolumn{1}{|c|}{\multirow{2}{*}{2.7}} & \multirow{2}{*}{42.7}  &{\small  +1.57\% +0.59\%} & \multicolumn{1}{|c}{\multirow{2}{*}{41.5}}  &\multicolumn{1}{c|} {\small +1.61\% +0.63\% } \\ \cline{9-9} \cline{11-11} 
      \multicolumn{1}{|c|}{}             &                   &       \multicolumn{1}{|c|}{}              &                   &             \multicolumn{1}{|c|}{}          &                   &          \multicolumn{1}{|c|}{}             &                    &{\small -1.29\% -0.58\%}  &             \multicolumn{1}{|c}{}           &\multicolumn{1}{c|} {\small -1.40\% -0.60\%}  \\ \hline
\end{tabular}
\end{table}

\begin{multicols}{2}

\section{Limits on compositeness scale}
If the $\mu_{8}$ is discovered by the FCC-pp option, $\mu p$ colliders
will give opportunity to estimate compositeness scale. In this regard,
two distinct possibilities should be considered: 

a)$\;$$\mu_{8}$ is discovered by the FCC but not observed at $\mu$-FCC.
In this case one can put lower limit on compositeness scale, 

b)$\;$$\mu_{8}$ is discovered by the FCC and also observed at $\mu$-FCC.
In this case one can determine compositeness scale. 

In this section we present the analysis of these two possibilities
for four different benchmark points, namely, $M_{\mu_{8}}=2.5,\,5,\,7.5$
and $10$ TeV.

\subsection{$\mu_{8}$ is discovered by the FCC but not observed at $\mu$-FCC}

If we assume that $\mu_{8}$ mass is found out by FCC results then
it is possible to determine optimal cuts for given $M_{\mu_{8}}$
at the $\mu$-FCC colliders. Let us start by consideration of $M_{\mu_{8}}=5.0$
TeV at $\mu$750$\otimes$FCC.

It is seen from Fig. 8 that $-1.3<\eta_{\mu}<4.74$ and $0.7<\eta_{j}<3.3$
cuts drastically decrease the background whereas the signal is slightly
affected. Similar cuts were determined for other $\mu$-FCC collider
options and $M_{\mu_{8}}$ values and these optimal cuts were presented
in Table III. Invariant mass window $0.99M_{\mu_{8}}<M_{\mu j}<1.01M_{\mu_{8}}$
has been used in this particular analysis. $\mu$63$\otimes$FCC
collider was not included in this section due to its remarkably low
potential compared to the other options.

\end{multicols}
\begin{center}
\includegraphics[width=18cm]{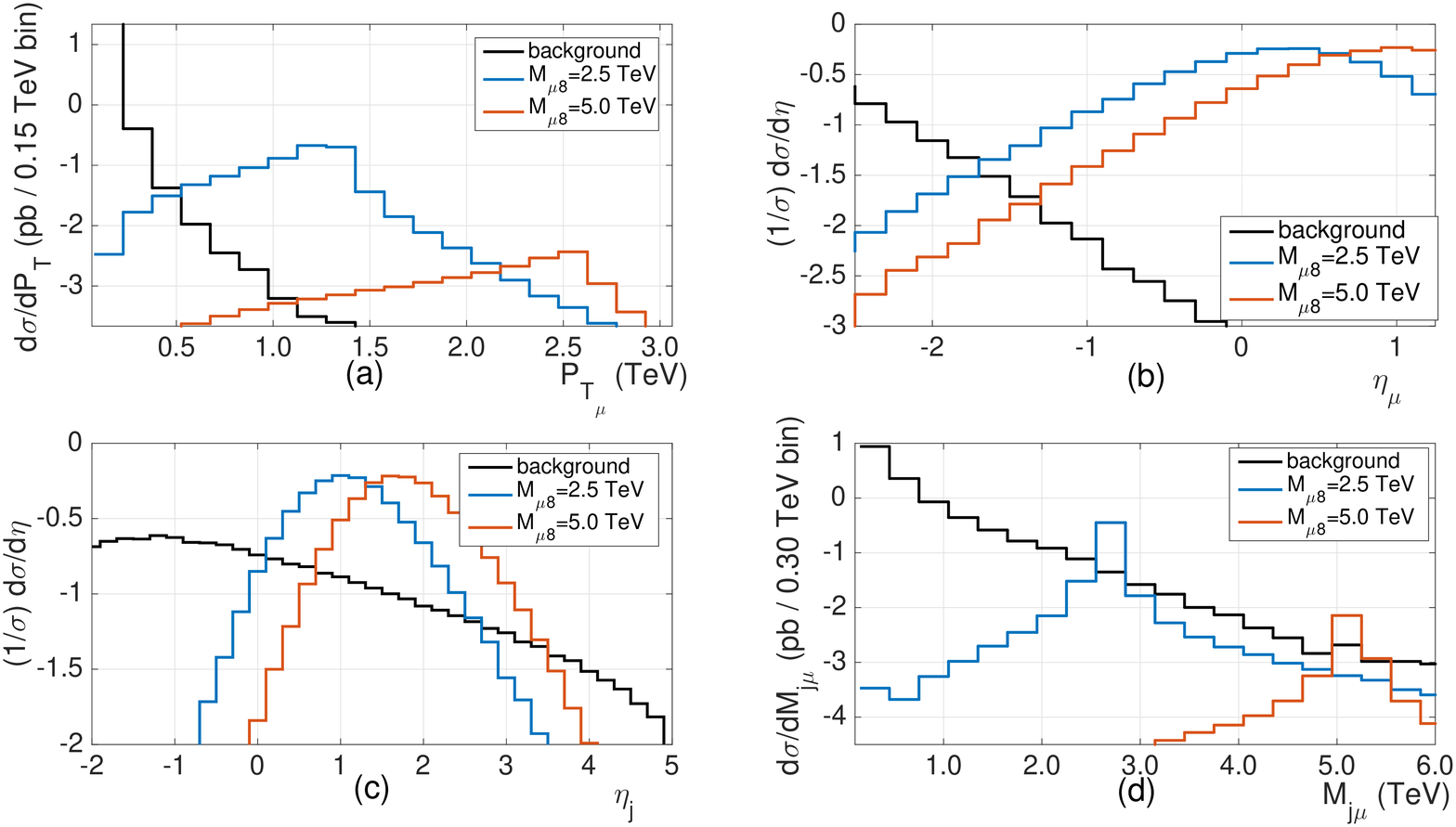}
\figcaption{\label{fig8} a) Transverse momentum distributions of final state muons (almost same distribution holds for the leading-jets),
b) pseudorapidity distributions of final state muons, c) pseudorapidity
distributions of final state jets and d) invariant mass distributions
for signal and background at $\mu750\otimes$FCC after generic cuts. }
\end{center}
\begin{multicols}{2}

\end{multicols}
\begin{table}

\tabcaption{\label{tab3} Optimal cuts for determination of compositeness scale lower bounds.}

\centering{}%
\begin{tabular}{|c|c|c|c|c|c|c|c|c|c|}
\hline 
\multirow{2}{*}{Collider} & \multirow{2}{*}{Cut Type} & \multicolumn{2}{c|}{$M_{\mu_{8}}=2.5$ TeV} & \multicolumn{2}{c|}{$M_{\mu_{8}}=5.0$ TeV} & \multicolumn{2}{c|}{$M_{\mu_{8}}=7.5$ TeV} & \multicolumn{2}{c|}{$M_{\mu_{8}}=10$ TeV}\tabularnewline
\cline{3-10} 
 &  & min & max & min & max & min & max & min & max\tabularnewline
\hline 
\multirow{3}{*}{{$\mu$750$\otimes$FCC}} & $\eta_{\mu}$ & -1.7 & 4.74 & -1.3 & 4.74 & -1.2 & 4.74 & - & -\tabularnewline
\cline{2-10} 
 & $\eta_{j}$ & 0.2 & 2.6 & 0.7 & 3.3 & 1.0 & 3.9 & - & -\tabularnewline
\cline{2-10} 
 & Mass Window (GeV) & 2475 & 2525 & 4950 & 5050 & 7425 & 7575 & - & -\tabularnewline
\hline 
\multirow{3}{*}{{$\mu$1500$\otimes$FCC}} & $\eta_{\mu}$ & -2.3 & 4.74 & -2.0 & 4.74 & -1.8 & 4.74 & -1.7 & 4.74\tabularnewline
\cline{2-10} 
 & $\eta_{j}$ & -0.6 & 1.9 & -0.1 & 2.7 & 0.4 & 3.1 & 0.5 & 3.5\tabularnewline
\cline{2-10} 
 & Mass Window (GeV) & 2475 & 2525 & 4950 & 5050 & 7425 & 7575 & 9900 & 10100\tabularnewline
\hline 
\multirow{3}{*}{{$\mu$3000$\otimes$FCC}} & $\eta_{\mu}$ & -2.9 & 4.74 & -2.7 & 4.74 & -2.5 & 4.74 & -2.3 & 4.74\tabularnewline
\cline{2-10} 
 & $\eta_{j}$ & -1.4 & 1.4 & -0.8 & 2.1 & -0.4 & 2.6 & -0.2 & 3.1\tabularnewline
\cline{2-10} 
 & Mass Window (GeV) & 2475 & 2525 & 4950 & 5050 & 7425 & 7575 & 9900 & 10100\tabularnewline
\hline 
\multirow{3}{*}{{$\mu$20000$\otimes$FCC}} & $\eta_{\mu}$ & -3.9 & 4.74 & -3.5 & 4.74 & -3.3 & 4.74 & -3.2 & 4.74\tabularnewline
\cline{2-10} 
 & $\eta_{j}$ & -3.0 & -0.9 & -2.5 & 0.1 & -2.1 & 0.5 & -1.9 & 1.0\tabularnewline
\cline{2-10} 
 & Mass Window (GeV) & 2475 & 2525 & 4950 & 5050 & 7425 & 7575 & 9900 & 10100\tabularnewline
\hline 
\end{tabular}
\end{table}
\begin{multicols}{2}

Applying cuts presented in Table III and $p_{T}>350$ GeV for all
cases one can estimate achievable lower limits on compositeness scale.
Using Eq. 5 we obtain $\Lambda$ values given in Table IV. As expected,
lower bounds on compositeness scale is decreased with increasing value
of the $\mu_{8}$ mass. It is seen that multi-hundred TeV lower bounds
can be put on the compositeness scale if $\mu_{8}$ is discovered at the
FCC and not observed at any $\mu$ $\otimes$ FCC.

\end{multicols}
\begin{table}
\tabcaption{\label{tab4} Lower limits on compositeness scale in TeV units at the FCC based
$\mu p$ colliders.}

\centering{}%
\begin{tabular}{|c|c|c|c|c|c|c|c|c|c|}
\hline 
\multirow{2}{*}{Collider} & \multirow{2}{*}{$L_{int}$, $fb^{-1}$} & \multicolumn{2}{c|}{$M_{\mu_{8}}=2.5$ TeV} & \multicolumn{2}{c|}{$M_{\mu_{8}}=5.0$ TeV} & \multicolumn{2}{c|}{$M_{\mu_{8}}=7.5$ TeV} & \multicolumn{2}{c|}{$M_{\mu_{8}}=10$ TeV}\tabularnewline
\cline{3-10} 
 &  & $3\sigma$ & $5\sigma$ & $3\sigma$ & $5\sigma$ & $3\sigma$ & $5\sigma$ & $3\sigma$ & $5\sigma$\tabularnewline
\hline 
{$\mu$750$\otimes$FCC} & 5 & 270 & 210 & 170 & 130 & 50 & 35 & - & -\tabularnewline
\hline 
{$\mu$1500$\otimes$FCC} & 5 & 360 & 280 & 220 & 170 & 130 & 100 & 55 & 40\tabularnewline
\hline 
{$\mu$3000$\otimes$FCC} & 5 & 475 & 370 & 320 & 245 & 230 & 170 & 140 & 105\tabularnewline
\hline 
{$\mu$20000$\otimes$FCC} & 10 & 1390 & 1080 & 850 & 655 & 515 & 400 & 315 & 246\tabularnewline
\hline 
\end{tabular}
\end{table}
\begin{multicols}{2}

\subsection{$\mu_{8}$ is discovered by the FCC and observed at $\mu$-FCC}

In this case, the value of cross section at $\mu p$ colliders, which
is inversely proportional to $\varLambda^{2}$ gives the opportunity to
determine the compositeness scale directly. As an example, let us consider the $\mu1500\otimes$FCC case. In Fig. 9 we present $\varLambda$ dependence
of $\mu_{8}$ production cross section for $M_{\mu_{8}}=2.5,\,5,\,7.5$
TeV. Supposing that FCC discovers $\mu_{8}$ with 5 TeV mass and $\mu1500$-FCC
measures cross section as $\sigma_{exp}\sim100$ $fb$, one can derive the
compositeness scale as $\varLambda_{exp}\backsimeq70$ TeV.

\end{multicols}
\begin{center}
\includegraphics[width=18cm]{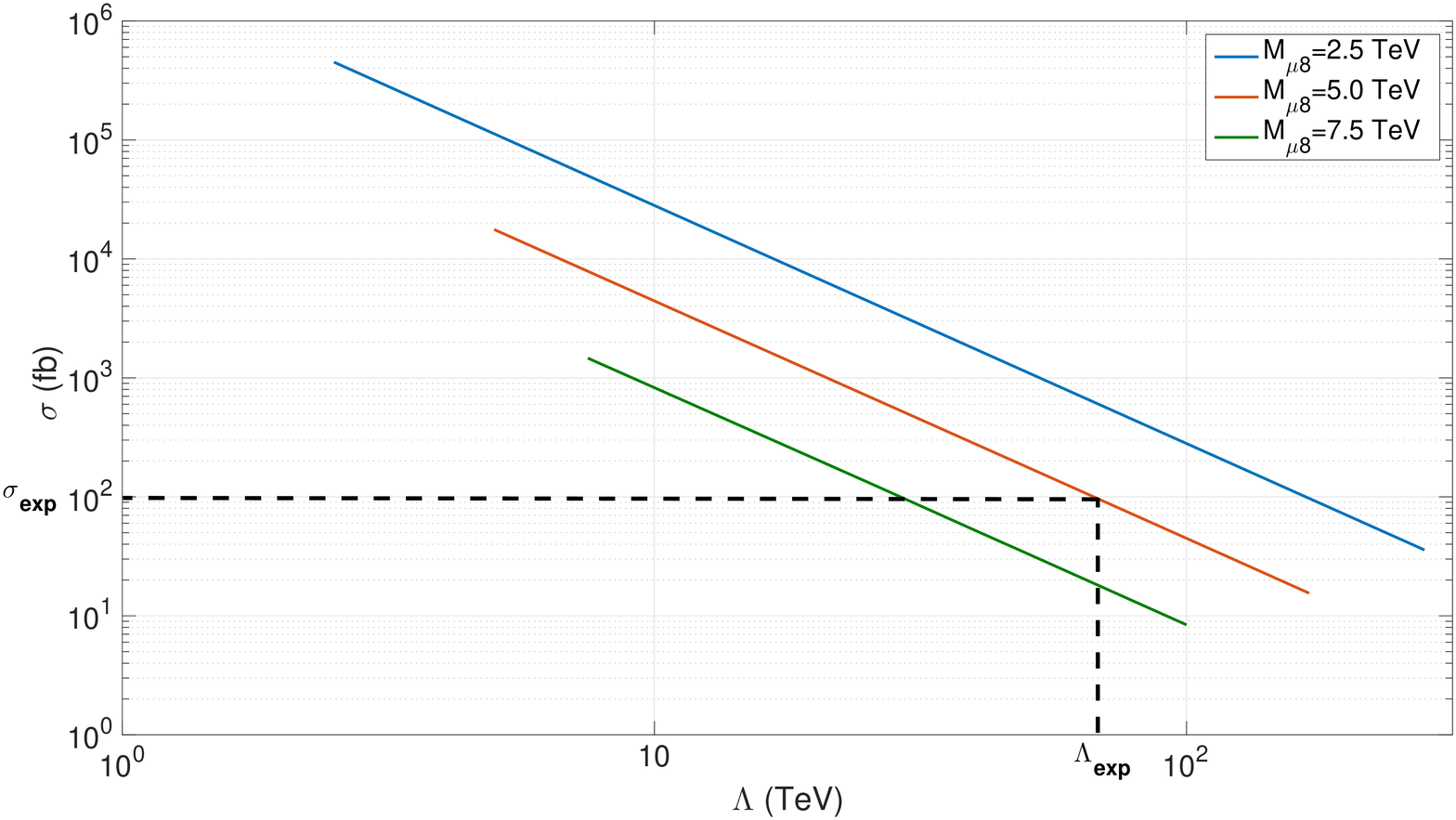}
\figcaption{\label{fig9}  Cross section distributions with respect to compositeness scale for
$\mu1500\otimes$FCC collider.}
\end{center}
\begin{multicols}{2}

\subsection{$\mu_{8}$ is not discovered by FCC but observed at $\mu$-FCC}

Another possibility is the failure of $\mu_{8}$ search at the FCC. This can be caused by the value of color-octet muon mass,  $M_{\mu8}$, which can be greater than the discovery limit of the FCC itself. In this case, the advantages of $\mu$-FCC colliders with quite large discovery limits manifest themselves.

\section{Conclusion}

Discovery mass limits for $\mu_{8}$ at the muon, proton and FCC based
$\mu p$ colliders are shown in Fig. 10. It is obvious
that discovery mass limits for pair production of $\mu_{8}$ at muon
colliders are approximately half of CM energies. Discovery limit
values for LHC and FCC are obtained by rescaling ATLAS/CMS second generation
LQ results \cite{lab16,lab17} using the method developed by G.
Salam and A. Weiler \cite{lab20}. Following \cite{lab21}, integrated luminosity values 3 ab$^{-1}$ and 20 ab$^{-1}$ have been used for the High Luminosity LHC (HL-LHC) and the FCC-hh, respectively. As can be seen from Fig. 10, FCC based $\mu p$ colliders with a discovery limit up to 40 TeV are the
most advantageous among the other collider options for $\mu_{8}$
searches. Moreover, FCC based $\mu p$ colliders will give the opportunity
to probe compositeness up to the PeV scale.

\acknowledgments {The authors are grateful to Saleh Sultansoy for useful discussions. The authors are
also grateful to Subhadip Mitra and Tanumoy Mandal for sharing their leptogluon
MadGraph model file. This paper is to be published in Chinese Physics C.}

\end{multicols}
\begin{center}
\includegraphics[width=18cm]{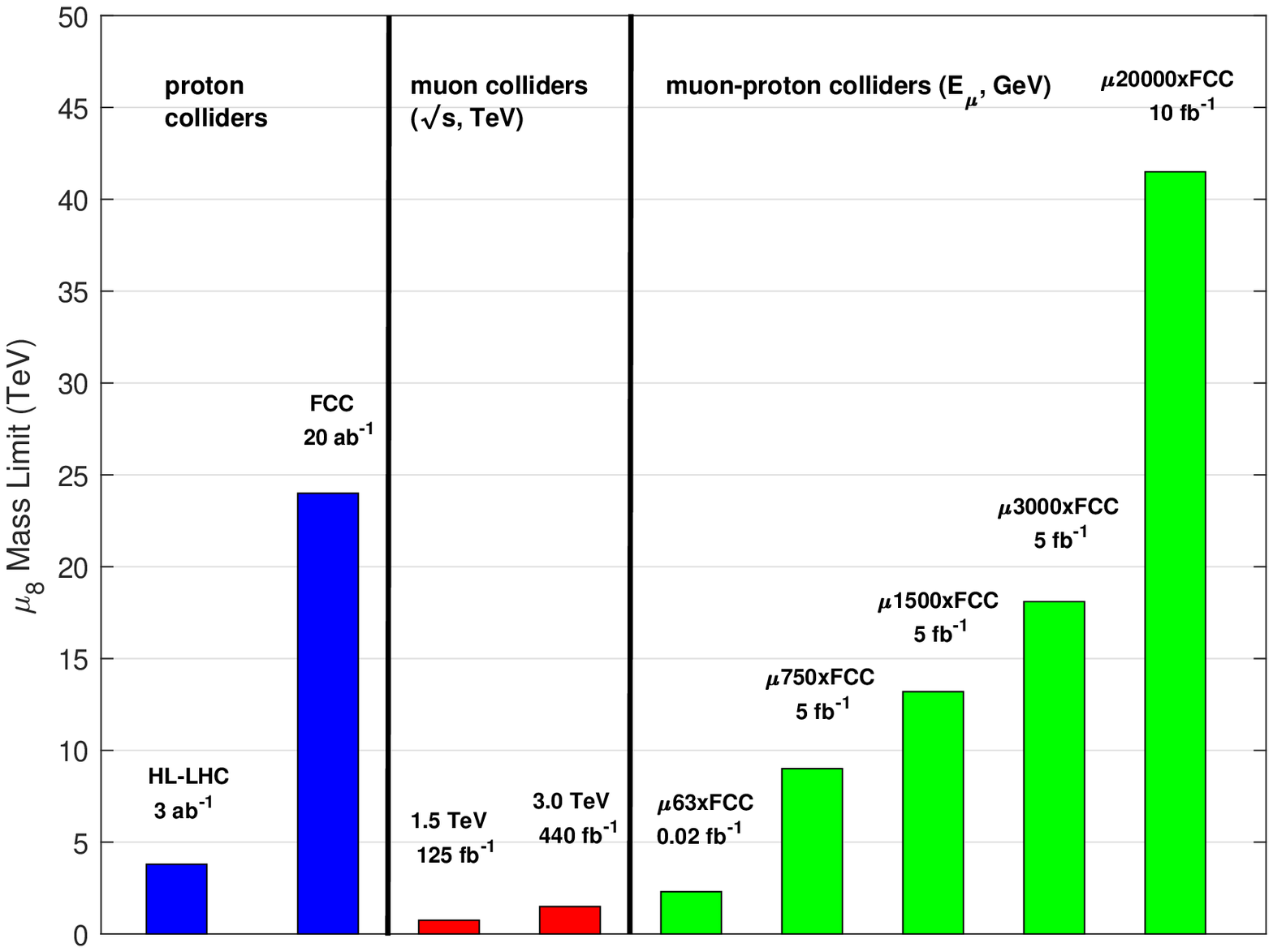}
\figcaption{\label{fig10}  Mass discovery limits ($SS=5$) of the color octet muon regarding
different type of colliders, i.e. proton, muon and muon-proton.}
\end{center}
\begin{multicols}{2}

\end{multicols}

\clearpage
\end{CJK*}
\end{document}